\documentclass[aps,prl,twocolumn]{revtex4}
\usepackage{graphicx}
\usepackage{epsfig}

\begin{document}

\title{Spontaneous currents and half-integer Shapiro steps in Superconductor-Ferromagnet-Superconductor 0-$\pi$ Josephson junctions}

\author{S. M. Frolov}
\author{D. J. Van Harlingen}

\affiliation{Department of Physics, University of Illinois at Urbana-Champaign, Urbana, IL, 61801,
USA}

\author{V. V. Bolginov}
\author{V. A. Oboznov}
\author{V. V. Ryazanov}

\affiliation{Institute of Solid State Physics, Russian Academy of Sciences, Chernogolovka, 142432,
Russia}

\date{\today}

\begin{abstract}

\noindent We study Superconductor-Ferromagnet-Superconductor (Nb-Cu$_{0.47}$Ni$_{0.53}$-Nb)
Josephson junctions with spatial variations in the barrier thickness. Critical current vs. magnetic
flux diffraction patterns indicate that the critical current density changes sign along the width
of the junctions, creating interfaces between 0 and $\pi$ junction regions around which spontaneous
currents can circulate. Shapiro steps observed at half-integer Josephson voltages can be explained
by the phase-locking of the spontaneous circulating currents to the applied rf modulation.
\end{abstract}


\maketitle In the past few years considerable attention has been
directed toward the understanding and realization of $\pi$ Josephson
junctions \cite{Bulaevskii}. Transitions between 0 and $\pi$
junction states were demonstrated in
Superconductor-Ferromagnet-Superconductor (SFS) junctions as a
function of temperature \cite{Ryazanov-junction} and barrier
thickness \cite{Aprili-junction}, and in mesoscopic
Superconductor-Normal metal-Superconductor (SNS) junctions as a
function of the barrier thermalization voltage
\cite{Klapwijk-junction}. In both systems, the first-order Josephson
supercurrent vanishes at the 0-$\pi$ transition, making the
transition region feasible for the observation of second-order
Josephson tunneling characterized by a $\sin(2\phi)$ component in
the current-phase relation (CPR). Period doubling in the critical
current $I_c$ vs. magnetic flux $\Phi$ patterns of dc SQUIDs
incorporating two mesoscopic SNS junctions, one biased at the
0-$\pi$ transition, has been attributed to a second-order Josephson
component \cite{Klapwijk-squid}. In SFS junctions, Shapiro steps at
half-integer Josephson voltages were reported close to a temperature
minimum in $I_c$, consistent with a $\sin(2\phi)$ Josephson
component \cite{Sellier-Sin2phi}. However, direct measurements of
the CPR in uniform SFS junctions revealed no discernible
$\sin(2\phi)$ term \cite{Frolov-CPR}.

In this paper, we present experimental evidence that half-integer Shapiro steps in the
current-voltage characteristics can occur in SFS junctions with a non-uniform critical current
density without the need for an intrinsic $\sin(2\phi)$ component. In our junctions, we observe
half-integer Shapiro steps, but only in a narrow temperature range close to the temperature at
which a deep but finite minimum occurs in the zero-field critical current. In this regime, we find
that part of the junction is in the 0 state and part is in the $\pi$ state with the net critical
currents of the two regions comparable in magnitude, and that the energy of the junction is
minimized by generation of a spontaneous circulating current. This current can couple to the
applied microwave field, producing half-integer Shapiro steps.  A similar effect has been observed
in an analogous system, a nearly-symmetric dc SQUID with an applied magnetic flux of
$\frac{1}{2}\Phi_0$ ($\Phi_0=h/2e$) \cite{Vanneste}.

A $\pi$ Josephson junction is characterized by a negative critical current. The mechanism of the
$\pi$ state in SFS Josephson junctions is the spatial oscillation of the proximity-induced order
parameter inside the ferromagnetic barrier which arises from its exchange field \cite{Buzdin}. The
critical current density $J_c$ is predicted to oscillate and decay with the barrier thickness $d$
according to \cite{Buzdin-review}

\begin{eqnarray}
J_c(d) \sim \left[\cos \left(\frac{d}{\xi_{F2}}\right) +\frac{\xi_{F1}}{\xi_{F2}}\sin
\left(\frac{d}{\xi_{F2}}\right)\right]\exp\left({-\frac{d}{\xi_{F1}}}\right) ,\label{eq:jc}
\end{eqnarray}

\noindent where $\xi_{F1}$ is the decay length and $2\pi\xi_{F2}$ is the oscillation period of the
order parameter.  This expression is valid for $d \gg \xi_{F1}$. The lengths $\xi_{F1}$ and
$\xi_{F2}$ can be extracted by fitting the measured $J_c(d)$ to Eq. (\ref{eq:jc}). In our junctions
at {T = 4.2 K}, {$\xi_{F1}$ $\approx$ 1.3 nm} and {$\xi_{F2}$ $\approx$ 3.7 nm}, and the first two
nodes of $J_c$ occur for {$d \approx$ 11 nm} and {$d \approx$ 22 nm}. The lengths $\xi_{F1}$ and
$\xi_{F2}$ vary with temperature according to

\begin{eqnarray}
\frac{\xi_{F1,F2}(T)}{\xi_{F1,F2}(0)} = \left\{ \frac{E_{ex}}{ \left[ (\pi k_B T)^2 + E_{ex}^2
\right]^{1/2} \pm \pi k_B T}\right\}^{1/2},\label{eq:ksi}
\end{eqnarray}

\noindent where $E_{ex}$ is the ferromagnet exchange energy and $\xi_{F1,F2}(0)$ are the values at
zero temperature. By using a weakly-ferromagnetic alloy {Cu$_{0.47}$Ni$_{0.53}$} with
$T_{Curie}\sim 60$ K as a barrier material, temperature changes in the 1-4 K range have a
significant effect on the suppression and modulation of the induced pair correlations in the
ferromagnetic interlayer, allowing us to tune through the 0-$\pi$ transition by changing the
temperature.

The SFS junctions were patterned by optical lithography.  Base and counterelectrode superconducting
layers were dc-sputtered Nb with thicknesses $100$ nm and $240$ nm respectively, separated by an
$11$ nm barrier layer of rf-sputtered {Cu$_{0.47}$Ni$_{0.53}$} and a 20-30 nm layer of dc-sputtered
Cu. The ferromagnetic layer thickness is chosen near the first 0-$\pi$ transition thickness, while
the Cu layer protects the barrier during processing. Junction sizes were 4 $\mu$m $\times$ 4 $\mu$m
or 10 $\mu$m $\times$ 10 $\mu$m, defined by a window in an insulating SiO layer deposited on top of
the CuNi/Cu barrier. Because SFS junctions have small $I_cR_N$ products $\sim$ 1-100 nV, a
commercial dc SQUID with a standard resistor $R_{st}\approx $ 10 m$\Omega$ is used as a
potentiometer with sensitivity 1 pV to perform transport measurements. This translates into a
critical current measurement resolution of 100 nA.  A uniform magnetic field up to 100 G can be
applied through the junction barrier from a solenoid coil to measure the magnetic field dependence
of the critical current, and an rf current component at frequencies $f_{rf}=$ 0.3-1.3 MHz is
superimposed with the dc bias current to observe ac-induced Shapiro steps in the current-voltage
characteristics.

\begin{figure}[b]
 \centering \includegraphics[width=7cm]{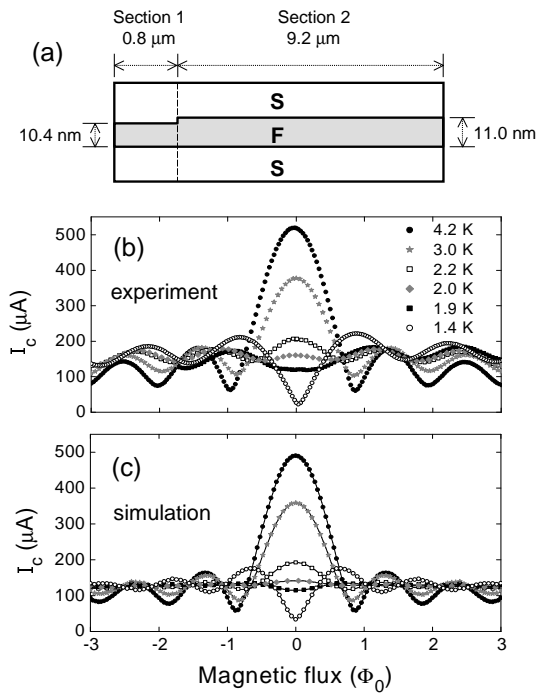}
 \caption{(a) Stepped ferromagnetic barrier deduced from critical current vs. applied magnetic flux measurements.
 (b) Diffraction patterns at a series of temperatures showing deviations from Fraunhofer behavior at low temperatures.
 (c) Simulated diffraction patterns using the deduced ferromagnet barrier profile. \label{diffraction}}
\end{figure}

Measurements of the critical current $I_c$ vs. the applied magnetic flux $\Phi$ threading the
junction barrier reveal that the critical current distribution is often not uniform across the
junction. Figure \ref{diffraction}(b) shows a series of $I_c(\Phi)$ curves in the temperature range
1.4-4.2 K for a 10 $\mu$m $\times$ 10 $\mu$m junction. At T = 4.2 K, $I_c(\Phi)$ has a
Fraunhofer-like shape but with non-vanishing supercurrents at the side minima. This can occur in a
junction with a localized region of high critical current density. In the temperature interval
1.4-1.9 K a minimum in the critical current is observed at zero field, indicating that regions of
opposite polarity critical current density exist in the junction.

\begin{figure}[t]
 \centering
 \includegraphics[width=6.5cm]{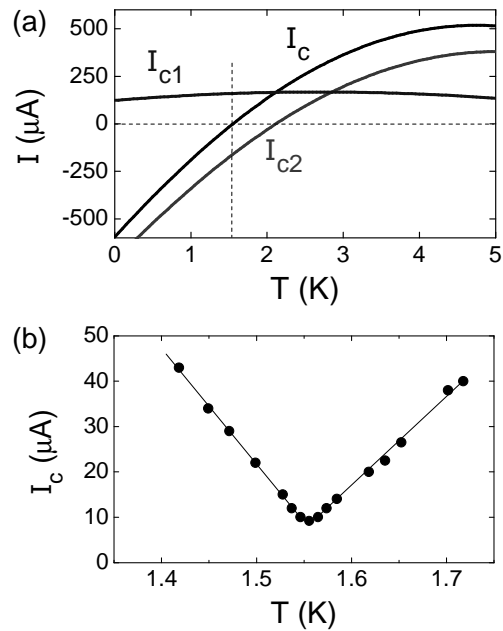}
 \caption{(a) Calculated temperature variation of the critical current for junction sections 1 and 2 and for the entire
junction. The thin region remains in the 0 state at all temperatures while the thick region crosses
from the 0 state into the $\pi$ state. (b) Measured critical current vs. temperature at zero
applied magnetic field dips but remains finite. \label{barrier}}
\end{figure}

The temperature evolution of the $I_c(\Phi)$ patterns indicates that some fraction of the junction
width makes a transition from the 0 state to the $\pi$ state as the temperature is lowered, while
the remaining part stays in the 0 state. The critical current non-uniformity likely arises from
spatial variations in the barrier thickness across the junction but could also be caused by
inhomogeneities in the ferromagnet exchange energy or by variations in the S-F interface
transparency. Figure \ref{diffraction}(a) shows the step barrier geometry deduced by fitting the
measured diffraction patterns in Fig. \ref{diffraction}(b) in the short junction approximation in
which magnetic fields from the tunneling current are neglected.  We obtain good agreement as
demonstrated in Fig. \ref{diffraction}(c). It is not surprising that the small 6 $\AA$ step has
such a dramatic effect on the diffraction patterns since close to the 0-$\pi$ transition the
critical current density in our junctions changes by 1000 A/cm$^2$ per 1 nm change in the barrier
thickness. For the barrier profile in Fig. \ref{diffraction}(a) and the experimental data for
$J_c(d)$, we use Eqs. (\ref{eq:jc}) and (\ref{eq:ksi}) to calculate the temperature dependences of
the zero-field critical currents $I_{c1}$ of the thin narrow region, $I_{c2}$ of the wide thick
region, and $I_{c}$, the total junction critical current. These are plotted in Fig.
\ref{barrier}(a). We see that $I_{c1}$ is relatively constant while $I_{c2}$ decreases and changes
sign at $T\approx$ 2.1 K, causing the $I_c$ to vanish at $T_{\pi0} \approx$ 1.55 K. However,
measurements plotted in Fig. \ref{barrier}(b) show that $I_c(\Phi=0)$ does not go fully to zero at
$T_{\pi0}$, instead reaching a minimum value of $\approx$ 10 $\mu$A.

\begin{figure}[b]
 \centering
 \includegraphics[width=6.5cm]{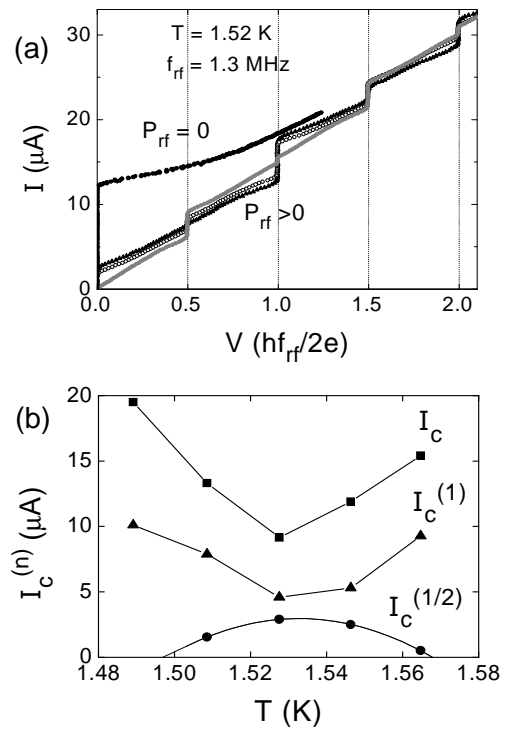}
 \caption{(a) Current-voltage characteristics showing rf-induced Shapiro steps both at the usual voltages
 $nhf_{rf}/2e$ and at half-integer values $nhf_{rf}/4e$. (b) Temperature dependence of the maximum (power-optimized)
 critical current steps, showing that the integer steps scale with the junction critical current whereas the
 half-integer steps only occur at temperatures near the minimum in the critical current. \label{fracShapiro}}
\end{figure}

Shapiro steps induced in zero applied magnetic field at drive frequency $f_{rf}$ are also
anomalous, exhibiting not only the usual steps at integer multiples of the Josephson voltage
$V_n=n(hf_{rf}/2e)$, but also steps at half-integer Josephson voltages such as $V_{1/2}=hf_{rf}/4e$
and $V_{3/2}=3hf_{rf}/4e$ when the temperature is close to $T_{\pi0}$.  An example is given in Fig.
\ref{fracShapiro}(a) for $f_{rf}$ = 1.3 MHz.  Figure \ref{fracShapiro}(b) shows the maximum value
of the (V=0) Josephson supercurrent, and the maximum amplitudes of the n=$\frac{1}{2}$ and the n=1
Shapiro steps obtained by adjusting the rf-amplitude (the $n^{th}$ step amplitudes vary with rf
voltage $V_{rf}$ according to the corresponding Bessel functions $J_n(2eV_{rf}/hf_{rf})$ as
expected).  At temperatures far from $T_{\pi0}$, only integer Shapiro steps were observed.
Half-integer steps appear near the minimum in the critical current in a temperature range of width
{$\sim$ 60 mK}. Note that the minimum in the critical current measured in Fig. \ref{fracShapiro}(b)
is about 25 mK lower than that in Fig. \ref{barrier}(b) as a result of room-temperature annealing
of the ferromagnetic barrier during the 3 days between when the measurements were made
\cite{Ryazanov-junction}. In other junctions in which the critical current does vanish at
$T_{\pi0}$, half-integer Shapiro steps were not observed at any temperatures.

The non-vanishing critical current at $T_{\pi0}$ and half-integer Shapiro steps could result from a
sin$(2\phi)$ component present in the current-phase relation near the 0-$\pi$ transition.  However,
we believe that they are more readily explained by self-field effects that must be taken into
account in finite width 0-$\pi$ junctions. At temperatures close to $T_{\pi0}$, spontaneous
currents circulate around interfaces between $0$ and $\pi$ regions to lower the total energy of the
system \cite{Bulaevskii-0pi,Kirtley-0pi}. These circulating currents generate magnetic flux through
the junction that prevent the total critical current from vanishing at any applied magnetic field.
In addition, they can resonate with an rf bias current applied at twice the Josephson frequency,
resulting in half-integer Shapiro steps. A detailed calculation of the spatial distribution of
spontaneous currents and the net critical current in this regime requires a self-consistent
solution of the Sine-Gordon equations in the junction. Computation of the amplitudes of the
half-integer Shapiro steps further requires numerical simulations of the junction phase dynamics.
However, we can understand the onset and effects of the spontaneous currents in a 0-$\pi$ junction
in analogy to a dc SQUID with finite geometric inductance.

We consider a dc SQUID with a 0 junction of critical current $I_{c0}>0$ and a $\pi$ junction of
critical current $I_{c\pi}<0$ in a loop of inductance $L$.  In such a SQUID, phase coherence around
the SQUID loop precludes both junctions being in their low energy states and the junction phases
$\phi_0$ and $\phi_\pi$ depend on the inductance parameter $\beta_L=2\pi L I_c/\Phi_0$, where
$I_c=(|I_{c0}|+|I_{c\pi}|)/2$, and the critical current asymmetry
$\alpha=(|I_{c0}|-|I_{c\pi}|)/2I_c$. For $\beta_L < 2\alpha/(1-\alpha^2)$, it is energetically
favorable to switch the phase of the junction with the smaller critical current magnitude into its
high energy state, {e. g.} if $|I_{c0}|>|I_{c\pi}|$, then $\phi_0$ = $\phi_\pi$ = 0. Under the
influence of a harmonic ac-drive, $\phi_0$ and $\phi_\pi$ wind in phase and only integer Shapiro
steps can be observed. When $\beta_L > 2\alpha/(1-\alpha^2)$, the SQUID energy is lowered by
generation of a spontaneous circulating current and $\phi_0 \neq \phi_\pi$. In this regime, the
phase differences $\phi_0$ and $\phi_\pi$ are no longer synchronized. Hence, the spontaneous
circulating current $J=(I_c/\beta_L)(\phi_0-\phi_\pi)$ can phase-lock to the driving frequency in
such a way that it switches direction an even or odd number of times during each period of the
drive signal, corresponding to integer or half-integer Shapiro steps. This phenomenon has
previously been studied in an equivalent system, an ordinary dc SQUID with an applied flux of
$\frac{1}{2} \Phi_0$. Measurements and simulations showed half-integer Shapiro steps with
amplitudes that increase as the SQUID inductance is increased and the critical current asymmetry is
decreased \cite{Vanneste}.

\begin{figure}[t]
 \centering
 \includegraphics[width=7cm]{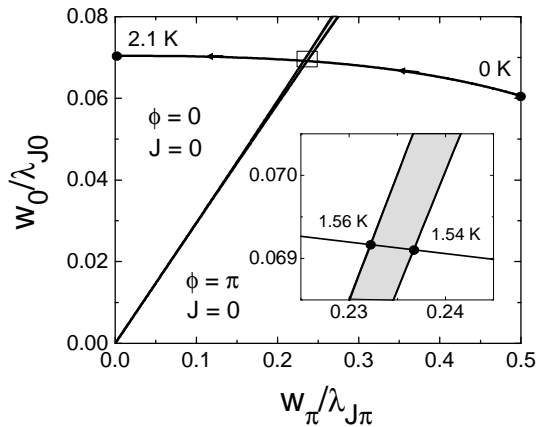}
 \caption{Phase diagram mapping the region of $w_0/\lambda_{J0}$ and $w_\pi/\lambda_{J\pi}$ values for which a
 spontaneous circulating current flows around the 0-$\pi$ step edge.  In the shaded region, the phase drop across the
 junction $\phi$ and current density $J$ vary spatially along the junction width.  Outside this region, the phase is
 uniform with value 0 or $\pi$ and no spontaneous currents flow in the junction. \label{vortex}}
\end{figure}

In a 0-$\pi$ junction, the existence of spontaneous circulating currents, and hence half-integer
Shapiro steps, depends on the ratio of the widths of the 0 and $\pi$ regions of the junction $w_0$
and $w_\pi$ to their corresponding Josephson penetration depths $\lambda_{J0}$ and $\lambda_{J\pi}$
\cite{Bulaevskii-0pi}. Here, $\lambda_J=(\hbar/2e \mu_0 d_m J_c)^{1/2}$, where the magnetic barrier
thickness $d_m=2\lambda+d$ depends on the penetration depth of the superconductor $\lambda$ and the
barrier thickness $d$. The phase diagram denoting the regimes of uniform phase and spontaneous
current flow is shown in Fig. \ref{vortex}. Regimes with uniform phase differences $\phi=0$ or
$\phi=\pi$ are separated by a narrow region in which the phase differences vary across the junction
due to spontaneous circulating currents. The circulating currents onset along lines defined by
$\lambda_{J\pi}\tanh(w_0/\lambda_{J0})=\lambda_{J0}\tan(w_\pi/\lambda_{J\pi})$ and
$\lambda_{J0}\tanh(w_\pi/\lambda_{J\pi})=\lambda_{J\pi}\tan(w_0/\lambda_{J0})$
\cite{Bulaevskii-0pi}. Also shown is the path in the phase diagram followed by the non-uniform
junction described in Fig. \ref{barrier}(a), for which $d_m$ = 100 nm, as the temperature is varied
from 0 K to 2.1 K, the temperature at which the wide segment of the junction crosses into the 0
state.  We see in the inset that spontaneous currents occur only in a narrow temperature range from
1.54-1.56 K.  This is comparable to the temperature range from 1.50-1.56 K in which we observe
half-integer Shapiro steps, suggesting that they have the same origin. Other non-uniform junctions
that we have measured either had smaller areas so that $w_0$ and $w_\pi$ were smaller, or smaller
barrier thickness steps so that $\lambda_{J0}$ and $\lambda_{J\pi}$ were larger.  In either case,
the temperature range of spontaneous currents was predicted to be immeasurably small ($<$ 1 mK),
explaining why we were not able to observe half-integer Shapiro steps or a finite minimum critical
current in these junctions.

Previously, we directly measured the current-phase relation of uniform SFS junctions with thicker
barriers $d \approx$ 22 nm, close to the second node of the $I_c(d)$ dependence, and did not
observe any signature of a sin($2\phi$) component in the Josephson tunneling \cite{Frolov-CPR}.
This is in agreement with microscopic predictions that the second-order Josephson tunneling
probability falls off more rapidly with increasing barrier thickness than the first-order Josephson
effect \cite{Buzdin-peculiar, Melin, Pistolesi}. Thus, to determine definitively whether or not a
$\sin(2\phi)$ component is present in the current-phase relation, uniform SFS Josephson junctions
with barrier thicknesses $d \approx$ 11 nm close to the first node of $I_c(d)$ should be studied.
Simultaneous observation of Shapiro steps at half-integer Josephson voltages and period doubling in
the current-phase relation in uniform SFS junctions is required to reliably verify the presence of
a microscopic sin$(2 \phi)$ component.

We thank Alexander Buzdin, Alexander Golubov, and Victor Vakaryuk for useful discussions. Work
supported by the National Science Foundation grant EIA-01-21568, by the U. S. Civilian Research and
Development Foundation (CRDF) grant RP1-2413-CG-02, and by the Russian Foundation for Basic
Research. We also acknowledge extensive use of the Microfabrication Facility of the Frederick Seitz
Materials Research Laboratory at the University of Illinois at Urbana-Champaign.

\bibliography{0-pi}

\begin{thebibliography}{15}
\expandafter\ifx\csname natexlab\endcsname\relax\def\natexlab#1{#1}\fi
\expandafter\ifx\csname bibnamefont\endcsname\relax
  \def\bibnamefont#1{#1}\fi
\expandafter\ifx\csname bibfnamefont\endcsname\relax
  \def\bibfnamefont#1{#1}\fi
\expandafter\ifx\csname citenamefont\endcsname\relax
  \def\citenamefont#1{#1}\fi
\expandafter\ifx\csname url\endcsname\relax
  \def\url#1{\texttt{#1}}\fi
\expandafter\ifx\csname urlprefix\endcsname\relax\def\urlprefix{URL }\fi
\providecommand{\bibinfo}[2]{#2}
\providecommand{\eprint}[2][]{\url{#2}}

\bibitem[{\citenamefont{Bulaevskii et~al.}(1977)\citenamefont{Bulaevskii,
  Kuzii, and Sobyanin}}]{Bulaevskii}
\bibinfo{author}{\bibfnamefont{L.~N.} \bibnamefont{Bulaevskii}},
  \bibinfo{author}{\bibfnamefont{V.~V.} \bibnamefont{Kuzii}}, \bibnamefont{and}
  \bibinfo{author}{\bibfnamefont{A.~A.} \bibnamefont{Sobyanin}},
  \bibinfo{journal}{JETP Lett.} \textbf{\bibinfo{volume}{25}},
  \bibinfo{pages}{290} (\bibinfo{year}{1977}).

\bibitem[{\citenamefont{Ryazanov et~al.}(2001)\citenamefont{Ryazanov, Oboznov,
  Rusanov, Veretennikov, Golubov, and Aarts}}]{Ryazanov-junction}
\bibinfo{author}{\bibfnamefont{V.~V.} \bibnamefont{Ryazanov}},
  \bibinfo{author}{\bibfnamefont{V.~A.} \bibnamefont{Oboznov}},
  \bibinfo{author}{\bibfnamefont{A.~Y.} \bibnamefont{Rusanov}},
  \bibinfo{author}{\bibfnamefont{A.~V.} \bibnamefont{Veretennikov}},
  \bibinfo{author}{\bibfnamefont{A.~A.} \bibnamefont{Golubov}},
  \bibnamefont{and} \bibinfo{author}{\bibfnamefont{J.}~\bibnamefont{Aarts}},
  \bibinfo{journal}{Phys. Rev. Lett.} \textbf{\bibinfo{volume}{86}},
  \bibinfo{pages}{2427} (\bibinfo{year}{2001}).

\bibitem[{\citenamefont{Kontos et~al.}(2002)\citenamefont{Kontos, Aprili,
  Lesueur, Genet, Stephanidis, and Boursier}}]{Aprili-junction}
\bibinfo{author}{\bibfnamefont{T.}~\bibnamefont{Kontos}},
  \bibinfo{author}{\bibfnamefont{M.}~\bibnamefont{Aprili}},
  \bibinfo{author}{\bibfnamefont{J.}~\bibnamefont{Lesueur}},
  \bibinfo{author}{\bibfnamefont{F.}~\bibnamefont{Genet}},
  \bibinfo{author}{\bibfnamefont{B.}~\bibnamefont{Stephanidis}},
  \bibnamefont{and} \bibinfo{author}{\bibfnamefont{R.}~\bibnamefont{Boursier}},
  \bibinfo{journal}{Phys. Rev. Lett.} \textbf{\bibinfo{volume}{89}},
  \bibinfo{pages}{173007} (\bibinfo{year}{2002}).

\bibitem[{\citenamefont{Baselmans et~al.}(1999)\citenamefont{Baselmans,
  Morpurgo, van Wees, and Klapwijk}}]{Klapwijk-junction}
\bibinfo{author}{\bibfnamefont{J.~J.~A.} \bibnamefont{Baselmans}},
  \bibinfo{author}{\bibfnamefont{A.~F.} \bibnamefont{Morpurgo}},
  \bibinfo{author}{\bibfnamefont{B.~J.} \bibnamefont{van Wees}},
  \bibnamefont{and} \bibinfo{author}{\bibfnamefont{T.~M.}
  \bibnamefont{Klapwijk}}, \bibinfo{journal}{Nature}
  \textbf{\bibinfo{volume}{397}}, \bibinfo{pages}{43} (\bibinfo{year}{1999}).

\bibitem[{\citenamefont{Baselmans et~al.}(2002)\citenamefont{Baselmans,
  Heikkil\"{a}, van Wees, and Klapwijk}}]{Klapwijk-squid}
\bibinfo{author}{\bibfnamefont{J.~J.~A.} \bibnamefont{Baselmans}},
  \bibinfo{author}{\bibfnamefont{T.~T.} \bibnamefont{Heikkil\"{a}}},
  \bibinfo{author}{\bibfnamefont{B.~J.} \bibnamefont{van Wees}},
  \bibnamefont{and} \bibinfo{author}{\bibfnamefont{T.~M.}
  \bibnamefont{Klapwijk}}, \bibinfo{journal}{Phys. Rev. Lett.}
  \textbf{\bibinfo{volume}{89}}, \bibinfo{pages}{207002}
  (\bibinfo{year}{2002}).

\bibitem[{\citenamefont{Sellier et~al.}(2004)\citenamefont{Sellier, Baraduc,
  Lefloch, and Calemczuk}}]{Sellier-Sin2phi}
\bibinfo{author}{\bibfnamefont{H.}~\bibnamefont{Sellier}},
  \bibinfo{author}{\bibfnamefont{C.}~\bibnamefont{Baraduc}},
  \bibinfo{author}{\bibfnamefont{F.}~\bibnamefont{Lefloch}}, \bibnamefont{and}
  \bibinfo{author}{\bibfnamefont{R.}~\bibnamefont{Calemczuk}},
  \bibinfo{journal}{Phys. Rev. Lett.} \textbf{\bibinfo{volume}{92}},
  \bibinfo{pages}{257005} (\bibinfo{year}{2004}).

\bibitem[{\citenamefont{Frolov et~al.}(2004)\citenamefont{Frolov, {Van
  Harlingen}, Oboznov, Bolginov, and Ryazanov}}]{Frolov-CPR}
\bibinfo{author}{\bibfnamefont{S.~M.} \bibnamefont{Frolov}},
  \bibinfo{author}{\bibfnamefont{D.~J.} \bibnamefont{{Van Harlingen}}},
  \bibinfo{author}{\bibfnamefont{V.~A.} \bibnamefont{Oboznov}},
  \bibinfo{author}{\bibfnamefont{V.~V.} \bibnamefont{Bolginov}},
  \bibnamefont{and} \bibinfo{author}{\bibfnamefont{V.~V.}
  \bibnamefont{Ryazanov}}, \bibinfo{journal}{Phys. Rev. B}
  \textbf{\bibinfo{volume}{70}}, \bibinfo{pages}{144505}
  (\bibinfo{year}{2004}).

\bibitem[{\citenamefont{Vanneste et~al.}(1988)\citenamefont{Vanneste, Chi,
  Gallagher, Kleinsasser, Raider, and Sandstrom}}]{Vanneste}
\bibinfo{author}{\bibfnamefont{C.}~\bibnamefont{Vanneste}},
  \bibinfo{author}{\bibfnamefont{C.~C.} \bibnamefont{Chi}},
  \bibinfo{author}{\bibfnamefont{W.~J.} \bibnamefont{Gallagher}},
  \bibinfo{author}{\bibfnamefont{A.~W.} \bibnamefont{Kleinsasser}},
  \bibinfo{author}{\bibfnamefont{S.~I.} \bibnamefont{Raider}},
  \bibnamefont{and} \bibinfo{author}{\bibfnamefont{R.~L.}
  \bibnamefont{Sandstrom}}, \bibinfo{journal}{J. Appl. Phys.}
  \textbf{\bibinfo{volume}{64}}, \bibinfo{pages}{242} (\bibinfo{year}{1988}).

\bibitem[{\citenamefont{Buzdin et~al.}(1982)\citenamefont{Buzdin, Bulaevskii,
  and Panjukov}}]{Buzdin}
\bibinfo{author}{\bibfnamefont{A.~I.} \bibnamefont{Buzdin}},
  \bibinfo{author}{\bibfnamefont{L.~N.} \bibnamefont{Bulaevskii}},
  \bibnamefont{and} \bibinfo{author}{\bibfnamefont{S.~V.}
  \bibnamefont{Panjukov}}, \bibinfo{journal}{JETP Lett.}
  \textbf{\bibinfo{volume}{35}}, \bibinfo{pages}{178} (\bibinfo{year}{1982}).

\bibitem[{\citenamefont{Buzdin}(2005{\natexlab{a}})}]{Buzdin-review}
\bibinfo{author}{\bibfnamefont{A.~I.} \bibnamefont{Buzdin}},
  \bibinfo{journal}{cond-mat/0505583}  (\bibinfo{year}{2005}{\natexlab{a}}).

\bibitem[{\citenamefont{Bulaevskii et~al.}(1978)\citenamefont{Bulaevskii,
  Kuzii, and Sobyanin}}]{Bulaevskii-0pi}
\bibinfo{author}{\bibfnamefont{L.~N.} \bibnamefont{Bulaevskii}},
  \bibinfo{author}{\bibfnamefont{V.~V.} \bibnamefont{Kuzii}}, \bibnamefont{and}
  \bibinfo{author}{\bibfnamefont{A.~A.} \bibnamefont{Sobyanin}},
  \bibinfo{journal}{Solid State Commun.} \textbf{\bibinfo{volume}{25}},
  \bibinfo{pages}{1053} (\bibinfo{year}{1978}).

\bibitem[{\citenamefont{Kirtley et~al.}(1997)\citenamefont{Kirtley, Moler, and
  Scalapino}}]{Kirtley-0pi}
\bibinfo{author}{\bibfnamefont{J.~R.} \bibnamefont{Kirtley}},
  \bibinfo{author}{\bibfnamefont{K.~A.} \bibnamefont{Moler}}, \bibnamefont{and}
  \bibinfo{author}{\bibfnamefont{D.~J.} \bibnamefont{Scalapino}},
  \bibinfo{journal}{Phys. Rev. B} \textbf{\bibinfo{volume}{56}},
  \bibinfo{pages}{886} (\bibinfo{year}{1997}).

\bibitem[{\citenamefont{Buzdin}(2005{\natexlab{b}})}]{Buzdin-peculiar}
\bibinfo{author}{\bibfnamefont{A.~I.} \bibnamefont{Buzdin}},
  \bibinfo{journal}{cond-mat/0505549}  (\bibinfo{year}{2005}{\natexlab{b}}).

\bibitem[{\citenamefont{M\'{e}lin}(2005)}]{Melin}
\bibinfo{author}{\bibfnamefont{R.}~\bibnamefont{M\'{e}lin}},
  \bibinfo{journal}{Europhys. Lett.} \textbf{\bibinfo{volume}{69}},
  \bibinfo{pages}{121} (\bibinfo{year}{2005}).

\bibitem[{\citenamefont{Houzet et~al.}(2005)\citenamefont{Houzet, Vinokur, and
  Pistolesi}}]{Pistolesi}
\bibinfo{author}{\bibfnamefont{M.}~\bibnamefont{Houzet}},
  \bibinfo{author}{\bibfnamefont{V.}~\bibnamefont{Vinokur}}, \bibnamefont{and}
  \bibinfo{author}{\bibfnamefont{F.}~\bibnamefont{Pistolesi}},
  \bibinfo{journal}{cond-mat/0505514}  (\bibinfo{year}{2005}).

\end{thebibliography}

\end{document}